\documentclass{article}
\usepackage{stywhispers,amsmath,epsfig,graphicx,caption,subcaption,amssymb}
\usepackage[export]{adjustbox}
\usepackage{xcolor, indentfirst, float}
\usepackage{microtype}

\title{Unsupervised Super-Resolution of Hyperspectral remote sensing images using fully synthetic training}
\name{Xinxin Xu\thanks{The work was partially supported by Agence de l’Innovation de Défense – AID - via Centre Interdisciplinaire d’Etudes pour la Défense et la Sécurité – CIEDS - (project 2023 - ALIA).}, Yann Gousseau, Christophe Kervazo, Saïd Ladjal}
\address{Télécom Paris - Institut Polytechnique de Paris\\
	Department IDS\\
	19 Pl. Marguerite Perey, 91120 Palaiseau, France}
\begin{document}
%\ninept
%
\sloppy
\maketitle
\begin{abstract}
\indent Considerable work has been dedicated to hyperspectral single image super-resolution to improve the spatial resolution of hyperspectral images and fully exploit their potential. However, most of these methods are supervised and require some data with ground truth for training, which is often non-available. To overcome this problem, we propose a new unsupervised training strategy for the super-resolution of hyperspectral remote sensing images, based on the use of synthetic abundance data. Its first step decomposes the hyperspectral image into abundances and endmembers by unmixing. Then, an abundance super-resolution neural network is trained using synthetic abundances, which are generated using the dead leaves model in such a way as to faithfully mimic real abundance statistics. %This network is then used to super-resolve the considered hyperspectral image. 
Next, the spatial resolution of the considered hyperspectral image abundances is increased using this trained network, %and the super-resolved abundances are recombined with the image endmembers to obtain 
and the high resolution hyperspectral image is finally obtained by recombination with the endmembers. 
Experimental results show the training potential of the synthetic images, and demonstrate the method effectiveness.
\end{abstract}

\section{Introduction}
\label{sec:intro}

In contrast to usual multispectral images, hyperspectral images (HSI) acquire a scene in many contiguous and narrow spectral bands. Consequently, HSIs are 3-dimensional images having two spatial dimensions and one spectral dimension. 
Due to their rich spectral information, HSIs have many applications in remote sensing, such as land cover and land use~\cite{bounouh2017prediction}, geology, vegetation monitoring~\cite{mills2010evaluation} and astrophysics \cite{fahes2022unrolling} to only cite a few. However, due to optical constraints, hyperspectral sensors generally have larger pixels than their multispectral counterparts, leading to images having lower spatial resolution. %land cover and land use , geology \cite{zhizhong2012review}, vegetation \cite{mills2010evaluation}

In this context, HSI super-resolution, which aims at increasing the HSI spatial resolution while maintaining its spectral resolution, has received a broad interest in the last decades. Existing methods can be categorized into two categories: multi/hyperspectral fusion and Single Image Super-Resolution (SISR). 

The principle of multi/hyperspectral fusion is to merge a mutispectral image, having a good spatial resolution, with a hyperspectral image, having a good spectral resolution. Many fusion methods rely on Hyperspectral Unmixing, enabling to decompose the considered images into endmembers and abundances. The main insight is to use the abundances of thee multispectral image, containing most of the scene spatial information, in conjunction with the endmembers of the HSI, containing the spectral information. Building on this principle and on modern deep learning methods, Zheng et al. proposed HyCoNet, which structure is made of 3 autoencoders, and where the point spread function (PSF) and spectral response function (SRF) can be directly estimated from the images \cite{zheng2020coupled}. More recently, Hong et al. \cite{hong2023decoupled} have proposed a fusion network based on unmixing, with a preliminary step of decoupling the images into common and sensor-specific components before reuniting them in a new image space.

Nevertheless, while multi/hyperspectral fusion methods have received much attention, their use is impeded by the difficulty of perfectly co-registering the multi and hyperspectral image pairs. Consequently, we rather focus in this work on SISR methods. Early methods relied on model-based approaches. For instance, in 2005, Akgun et al.\cite{akgun2005super} formulated the super-resolution task as an inverse problem and used a method based on Projection Onto Convex Sets \cite{bauschke1996projection} to solve it. More recently data-driven methods have taken the lion share and numerous Convolutional Neural Networks (CNN) structures have been proposed. For instance, Yuan et al. \cite{yuan2017hyperspectral} used a transfer learning technique with a network trained on natural images. Other works have explored the use of three-dimensional convolutions to extract both spatial and spectral information. Typically, architectures can either use 3D-convolutions \cite{mei2017hyperspectral}, or a combination of 1D and 2D convolutions \cite{li2019dual}, or mix 2D and 3D convolutions \cite{li2020mixed, wang2020hyperspectral}.

\begin{figure*}[!htb]
    \centering
    \includegraphics[width=0.745\linewidth]{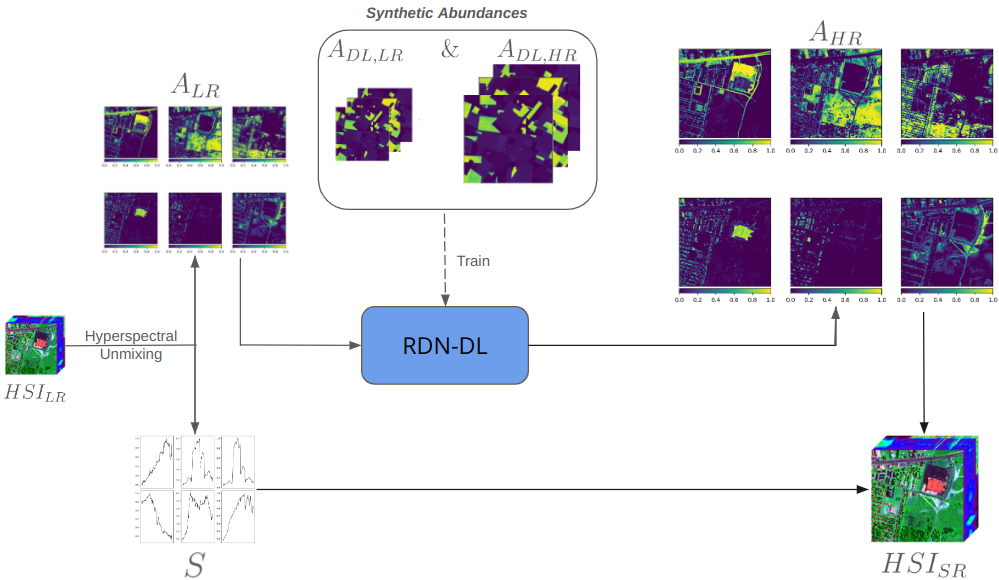}
    \caption{Structure of RDN-DL}
    \label{fig:Structure_DLSRnet}
\end{figure*}

Current SISR methods are however generally impeded by the lack of training sets: there are very few datasets, if any, containting the ground-truth high resolution image, which strongly limits the use of supervised methods. As such, we propose in this work a new method to generate synthetic HSIs to train super-resolution neural networks.

%In the context of natural images, synthetic image generation has been the subject of significant advances, with models such as Markov random fields \cite{cross1983markov}, wavelet models \cite{heeger1995pyramid}, Gaussian models \cite{galerne2011micro} or Dead Leaves model \cite{alvarez1999size, cao2010dead, gousseau2003dead}.

In order to be useful for the training of super-resolution networks, synthetic images should contain basic image structures such as edges and homogeneous regions. We choose to rely on the dead leaves model, which can produce such structures from  %model the non-Gaussianity of most observed statistics
only a few parameters and has been shown to be a powerful model for the training of neural networks in the case of natural images~\cite{achddou2023fully}. Consequently, we propose in this paper a new unsupervised SISR method based on a fully synthetic training using dead leaves model. The following section describes our method, and section \ref{sec: Results} shows results and ablation studies.

\section{Proposed method}
\label{sec:Proposed method}

\subsection{Overview}
\label{subsec:Overview}

To perform super-resolution, the objective is to obtain a high spatial resolution hyperspectral image $HSI_{HR} \in\mathbb{R}^{L \times H \times W}$ from a low resolution one, $HSI_{LR} \in\mathbb{R}^{L \times h \times w}$, where $h$, $w$, $H$ and $W$ ($h<H$ and $w<W$) are the spatial dimensions and $L$ is the spectral dimension. Due to its simplicity and despite its limitations \cite{kervazo2021provably}, we assume in this work that each pixel in $HSI$ can be exactly decomposed using the linear hyperspectral mixing model by:
\begin{equation}
    HSI_{LR}(l,i,j) = \sum_{n=1}^{N} S(l,n) \cdot A_{LR}(n,i,j)
    \label{eq:unmixing}
\end{equation}
where $A_{LR} \in \mathbb{R}^{N \times h \times w}$ and $S \in \mathbb{R}^{L \times N}$ are the abundances maps and the endmembers, respectively, of the low resolution HSI, and $N$ is the number of endmembers (to be fixed by the user). In this work, we rule out the unmixing errors by assuming Formula~\eqref{eq:unmixing} to hold exactly (in the experimental section this is done using the unmixing ground truths provided with the considered datasets). The choice of the best unmixing method to use to estimate $A_{LR}$ and $S$ is left for future work.

From the above decomposition, our method main insight is to generate a large quantity $N_{train}$ of synthetic abundance maps $A_{DL,LR} \in \mathbb{R}^{N \times h \times w}$ and $A_{DL,HR} \in \mathbb{R}^{N \times H \times W}$  by using the dead leaves model. The high resolution images $A_{DL,HR}$ are first generated using this model, and the corresponding images $A_{DL,LR}$ are obtained using a PSF simulated with Gaussian blur and bicubic downsampling. The pairs $(A_{DL,HR},A_{DL,LR})$ are then used to train a super-resolution neural network, which is thus trained on synthetic abundances maps. Afterwards, at test time, the abundances $A_{LR}$ are given as input to obtain an estimate of the abundances at high resolution ${A}_{HR}$ . The final high-resolution image $HSI_{HR}$ is obtained by reconstructing $HSI_{HR}(l,i,j) = \sum_{n=1}^{N} S(l,n) \cdot A_{HR}(n,i,j)$. The structure of our method is summarized in Fig.~\ref{fig:Structure_DLSRnet}.

The following sections describe in detail the steps of the generation of synthetic abundances maps and the Super-Resolution CNN.

\subsection{Synthetic Abundance Generation using the Dead Leaves model}
\label{subsec:Synthetic Images using Dead Leaves model}
The dead leaves model was first introduced in 1968 by Matheron for modelling porous media \cite{matheron1968modele,  bordenave2006dead} and later proposed as a model for natural images~\cite{alvarez1999size, lee2001occlusion, gousseau2007modeling}. The main idea is to generate an image by sequentially superimposing random shapes at random positions, mimicking the process of dead leaves falling from a tree. The shapes can be defined using any random model, and their positions are given by a stationary Poisson point process (points uniformly spread over the plane). The process is iterated until a stationary state is reached, which in practice can be obtained using perfect simulation techniques~\cite{kendall1999perfect} : each new shape is placed {\it below} the previous shapes, until the image is fully covered.  Of interest to us, Achddou et al. has recently shown the effectiveness of super-resolving natural images with a network trained only from dead leaves synthetic images \cite{achddou2023fully}.

In our method, we generate synthetic abundance maps based on the dead leaves model using the above procedure. % to synthesize abundances from remote sensing images. 
As we focused on the context of remote sensing urban HSIs, the use of rectangular leaves is natural. Precisely, each leaf is parameterized by the quadruplet $(a, b, \theta, V)$, with $a$, $b$ the size of the rectangular leave, $\theta$ the angle of rotation and $V \in [0, 1]$ a random value representing the value of the leaf. In addition, the synthetic abundances are also generated in such a way as to respect the Abundance Non-negative Constraint ($ANC$) and Abundance Sum-to-one Constraint ($ASC$). The $ANC$ constraint is naturally provided by non-negative abundance values inside each leaf, but the $ASC$ constraint is more complex to integrate into the falling leaf values. Therefore, it is rather ensured at the end of abundances generation using a pixelwise sum-to-one normalization.

\begin{figure}
    \centering
    \includegraphics[width=0.8\linewidth]{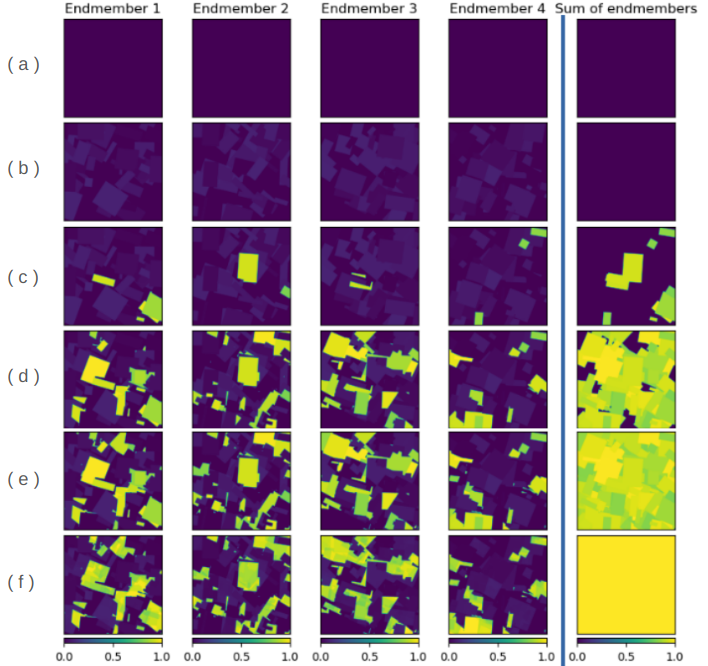}
    \caption{Example of Synthetic abundances generation (a) at the initialization (b) after the deposit of leaves to simulate local variations (c) after 10 leaves (d) after 100 leaves (e) at the end of the deposit process (f) after the $ASC$ constraint}
    \label{fig:DL_exmemple}
\end{figure}

In addition, in order to emulate the existing local variations in a given material abundance, a starting dead leaves field is initialized with a local variations layer : tens of random leaves, with low abundance values, are used to simulate a field that contains materials of low concentrations. The resulting  process of creating a synthetic abundance map can be summarized in three steps: local variation layer, dead leaves covering, ASC normalization. An example of abundance generation with 4 endmembers is shown in Fig. \ref{fig:DL_exmemple}, while Fig. 3 shows the interest of the local variation layer: the material textures are much more faithfully represented.
%with{\color{red} CK : à partir de là c'est trop redondant avec ce que tu as déjà écrit, peut-être que tu devrais plutôt mettre un pseudo-code} The process of leaves depositing then begins as decribed above. Each leaf is assigned a random position, shape and endmember. Previous leaves overlap futures leaves within the same endmember and with other endmembers. The process is repeated until the entire image is covered. Finally, the Sum-to-one operation is applied to the resulting abundance map to guarantee the $ASC$ constraint and finalize the creation of a synthetic abundance map. 
%{\color{red} Je ne comprends pas cette phrase} Fig. \ref{fig:compare_local} shows the impact of this local variation layer applied before the start of the process on the rendering of post-ASC synthetic abundances, by comparing to one without local variation layer and a ground true abundance. 

\begin{figure}
    \centering
    \includegraphics[width=0.8\linewidth]{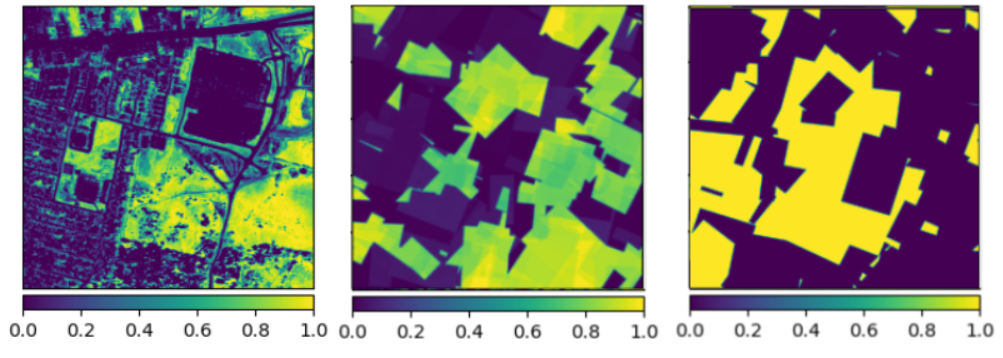}
    \caption{Comparison between a real abundance map of the \emph{Urban} dataset (Left); a synthetic Dead Leaves abundance map generated using the local variation layer as initialization (Middle); and a synthetic abundance map generated without the local variation layer (Right).}
    \label{fig:compare_local}
\end{figure}
%\vspace{-0.8cm}

\subsection{RDN-DL network}
\label{subsec : DLSRnet}

The Residual Dense Network (RDN) proposed by Zhang et al. \cite{zhang2018residual} is a high-performance supervised CNN  initially designed for the super-resolution of natural images. It is based on residual dense blocks for the local feature extraction and a global structure for dense feature fusion. RDN can be employed to process $K$-channels images. Consequently, it is possible to use this network to perform Super-Resolution on abundances by choosing $K = N$, the number of materials in the scene. To better adapt the RDN for abundance super-resolution, we further add an ASC layer at the end of the network structure, implemented as a softmax function.
% \begin{figure}[!htb]
%    \centering
%    \includegraphics[width=1\linewidth]{Figures/Ablation_fig.png}
%    \caption{Visual comparison of super-resolution results with a network trained fully with DLs, fully with Natural images and a Mixed ones}
%    \label{fig:fig_ablation}
%\end{figure}

\section{Experimental Results}
\label{sec: Results}

\subsection{Dataset \& Setup}
\label{subsec: Dataset}
The Urban dataset, a widely-used dataset where there is an exploitable ground-truth of the abundance maps and endmembers, is considered here. Urban is a dataset captured by the Hyperspectral Digital Image Collection Experiment (HYDICE) sensor. It is $307 \times 307$ pixels in size for $210$ bands in the $400 - 2500$ nm range. After removing noisy and corrupted bands, $162$ bands are usable. There are $N=6$ materials. As mentioned above, we do not directly work on Urban, but we rather use the available ground-truth of the abundances map $A_{HR}$ and endmembers $S$ \cite{zhu2017hyperspectral} to get rid of the hyperspectral unmixing errors. Precisely, all the compared methods are applied to: $HSI_{HR}(l,i,j) = \sum_{n=1}^{N} S(l,n) \cdot A_{HR}(n,i,j)$.

\begin{figure*}
    \centering
    \includegraphics[width=0.7\linewidth]{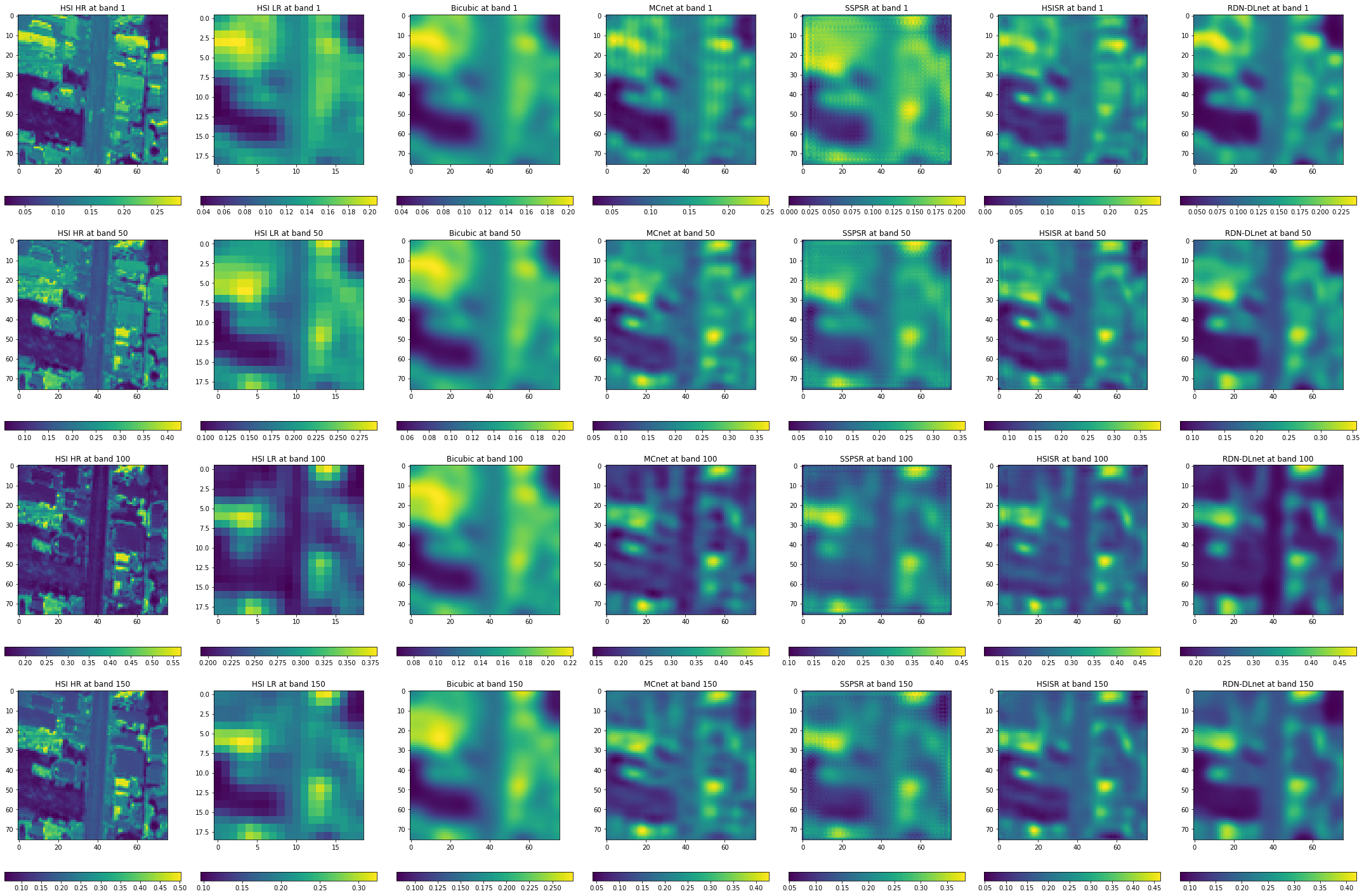}
    \caption{Visual comparison between the Ground Truth, LR, Bicubic, MCnet, SSPSR, HSISR and RDN-DL on one Urban's patch at the bande n°1, 50, 100 and 150}
    \label{fig:benchmark_fig}
\end{figure*}

To evaluate our algorithm performance, we work with a super-resolution factor of $4$. The $HSI_{LR}$ is obtained with a simulated PSF, by applying a Gaussian blur followed by a downsampling. The standard deviation of the Gaussian filter is fixed at $\sigma = 4$ to respect the  Nyquist–Shannon criterion, and the filter size is truncated at $6 \sigma$. For the training phase, we generate 10,000 synthetic $HR$ abundance maps of size $6 \times 500 \times 500$ following the protocol described above, and the $LR$ ones are obtained with the same PSF.

The network is trained for 100 epochs using L1 loss and the Adam optimizer, with a fixed learning rate set to 0.0001. We evaluate the results \cite{aburaed2023review} with Peak Signal-to-Noise Ratio (PSNR), Spectral Angle Mapper (SAM) and 
%Erreur Relative Globale Adimensionnelle de Synthèse 
Error relative global dimensionless synthesis (ERGAS).

\subsection{Comparison with the State-of-the-Art Methods}
\label{subsec: comparison}

In this section, we compare our method and the fully synthetic training dataset with other state-of-the-art SISR methods: MCnet \cite{li2020mixed}, SSPSR \cite{jiang2020learning}, and HSISR \cite{li2022hyperspectral}, we also add bicubic interpolation to our baseline. Since all these competitors are supervised methods, we need ground-truths to train them. To do that, we crop the HSI into 16 patches of size $162 \times 76 \times 76$. Then, 16 independent trainings of the methods are performed: each time, one of these 16 patches is reserved for performance evaluation, and the other 15 are used for training. Finally the 16 numerical values found per method and per metric are averaged.

This first shows the practical interest of our method: it is trained directly on the 10,000-large dataset of  dead leaves abundance maps, without any need of ground-truth. Table~\ref{tab:benchmark_tab} shows the quantitative results (computed over the whole HSI image), while Fig.~\ref{fig:benchmark_fig} shows the qualitative results for one patch. It can be seen that our method obtains very good results and outperforms the other methods. This is even more remarkable as the other methods would be difficult to apply in practice when no ground-truth is available.

\begin{table}
    \centering
    \caption{Average results between Bicubic, MCnet, SSPSR, HSISR and RDN-DL on 16 urban patches}
    \begin{tabular}{c|c|c|c|c|c}
         & Bicubic & MCnet  & SSPSR & HSISR & RDN-DL\\
         \hline
         mPSNR $\uparrow$& 26.485 & 27.475 & 26.380 & 27.549 &\textbf{27.784} \\
         mSAM $\downarrow$& 13.884 & 12.448 &13.666 & 12.329 & \textbf{12.138} \\
         mERGAS$\downarrow$& 7.328 & 6.578 & 7.264& 6.506 & \textbf{6.374} \\
    \end{tabular}
    \label{tab:benchmark_tab}
\end{table}

\section{Conclusion}
\label{Conclusion}

In this paper, we have proposed an unsupervised method for Super Resolution of remote sensing hyperspectral images. The main idea of our method is to train a supervised super-resolution network with synthetic dead leaves images only. This overcomes the problem of lack of datasets encountered by most super-resolution methods. Our method compares favorably with other state-of-the art algorithm. Future work will focus on the inclusion of the hyperspectral unmixing step in the overall super-resolution procedure.

\bibliographystyle{IEEEbib}
\bibliography{biblio}

\end{document}